\documentclass[11pt]{article}
\usepackage{amssymb}
\usepackage{amsfonts}
\usepackage{geometry}
\usepackage{mathrsfs}
\usepackage[onehalfspacing]{setspace}

\newtheorem{theorem}{Theorem}

\input{tcilatex}
\begin{document}

\title{Independent component models for replicated point processes}
\author{Daniel Gervini}
\maketitle

\begin{abstract}
We propose a semiparametric independent-component model for the intensity
functions of a point process. When independent replications of the process
are available, we show that the estimators are consistent and asymptotically
normal. We study the finite-sample behavior of the estimators by simulation,
and as an example of application we analyze the spatial distribution of
street robberies in the city of Chicago.

\end{abstract}

\section{Introduction}

Point processes in time and space have a broad range of applications, in
diverse areas such as neuroscience, ecology, finance, astronomy, seismology,
and many others. Examples are given in classic textbooks like Cox and Isham
(1980), Diggle (2013), M\o ller and Waagepetersen (2004), Streit (2010), and
Snyder and Miller (1991), and in the papers cited below. However, the
point-process literature has mostly focused on the single-realization case,
such as the distribution of trees in a single forest (Jalilian et al., 2013)
or the distribution of cells in a single tissue sample (Diggle et al.,
2006). But situations where several replications of a process are available
are increasingly common. Among the few papers proposing statistical methods
for replicated point processes we can cite Diggle et al.~(1991), Baddeley et
al.~(1993), Diggle et al.~(2000), Bell and Grunwald (2004), Landau et
al.~(2004), Wager et al.~(2004), and Pawlas (2011). However, these papers
mostly propose estimators for summary statistics of the processes (the
so-called $F$, $G$ and $K$ statistics) rather than for the intensity
functions that characterize the processes.

When several replications of a process are available, it is possible to
estimate the intensity functions directly thanks to the possibility of
\textquotedblleft borrowing strength\textquotedblright\ across replications.
This is the basic idea behind many functional data methods that are applied
to stochastic processes which, individually, are only sparsely sampled
(James et al., 2000; Yao et al., 2005; Gervini, 2009). Functional Data
Analysis has mostly focused on continuous-time processes; little work has
been done on discrete point processes. We can mention Bouzas et al.~(2006,
2007) and Fern\'{a}ndez-Alcal\'{a} et al.\ (2012), which have rather limited
scopes since they only estimate the mean of Cox processes, and Wu et
al.~(2013), who estimate the mean and the principal components of
independent and identically distributed realizations of a Cox process, but
the method of Wu et al.~(2013) is based on kernel estimators of the
covariance function that are not easily extended beyond i.i.d.~replications
and time-dependent processes. The semiparametric method we propose in this
paper, on the other hand, can be applied just as easily to temporal or
spatial processes, and can be extended to non-i.i.d.~situations like ANOVA
models or more complex data structures like marked point processes and
multivariate processes, even though we will not go as far in this paper.

As an example of application we will analyze the spatial distribution of
street robberies in Chicago in the year 2014, where each day is seen as a
replication of the process. The method we propose in this paper fits a
non-negative independent-component model for the intensity functions, and
provides estimators of the intensity functions for the individual
replications as by-products. We establish the consistency and asymptotic
normality of the component estimators in Section \ref{sec:Asymptotics},
study their finite-sample behavior by simulation in Section \ref%
{sec:Simulations}, and analyze the Chicago crime data in Section \ref%
{sec:Example}.

\section{Independent-component model for intensity process\label{sec:Model}}

A point process $X$ is a random countable set in a space $\mathcal{S}$,
where $\mathcal{S}$ is usually $\mathbb{R}$ for temporal processes and $%
\mathbb{R}^{2}$ or $\mathbb{R}^{3}$ for spatial processes (M\o ller and
Waagepetersen, 2004, ch.~2; Streit, 2010, ch.~2). Locally finite processes
are those for which $\#(X\cap B)<\infty $ with probability one for any
bounded $B\subseteq \mathcal{S}$. For such processes we can define the count
function $N(B)=\#(X\cap B)$. Given a locally integrable function $\lambda :%
\mathcal{S}\rightarrow \lbrack 0,\infty )$, i.e.~a function such that $%
\int_{B}\lambda (t)dt<\infty $ for any bounded $B\subseteq \mathcal{S}$, the
process $X$ is a Poisson process with intensity function $\lambda (t)$ if 
\emph{(i)} $N(B)$ follows a Poisson distribution with rate $\int_{B}\lambda
(t)dt$ for any bounded $B\subseteq \mathcal{S}$, and \emph{(ii)}
conditionally on $N(B)=m$, the $m$ points in $X\cap B$ are independent and
identically distributed with density $\tilde{\lambda}(t)=\lambda
(t)/\int_{B}\lambda $ for any bounded $B\subseteq \mathcal{S}$.

Let $X_{B}=X\cap B$ for a given $B$. Then the density function of $X_{B}$ at 
$x_{B}=\{t_{1},\ldots ,t_{m}\}$ is 
\begin{eqnarray}
f(x_{B}) &=&f(m)f(t_{1},\ldots ,t_{m}|m)  \label{eq:Pois_lik} \\
&=&\exp \left\{ -\int_{B}\lambda (t)dt\right\} \frac{\{\int_{B}\lambda
(t)dt\}^{m}}{m!}\times \prod_{j=1}^{m}\tilde{\lambda}(t_{j}).  \nonumber
\end{eqnarray}%
Since the realizations of $X_{B}$ are sets, not vectors, what we mean by\
`density' is the following: if $\mathcal{N}$ is the family of locally finite
subsets of $\mathcal{S}$, i.e. 
\[
\mathcal{N}=\left\{ A\subseteq \mathcal{S}:\#(A\cap B)<\infty \text{ for all
bounded }B\subseteq \mathcal{S}\right\} , 
\]%
then for any $F\subseteq \mathcal{N}$%
\begin{eqnarray*}
P\left( X_{B}\in F\right) &=&\sum_{m=0}^{\infty }\int_{B}\cdots \int_{B}%
\mathbb{I}(\{t_{1},\ldots ,t_{m}\}\in F)f(\{t_{1},\ldots
,t_{m}\})dt_{1}\cdots dt_{m} \\
&=&\sum_{m=0}^{\infty }\frac{\exp \left\{ -\int_{B}\lambda (t)dt\right\} }{m!%
}\int_{B}\cdots \int_{B}\mathbb{I}(\{t_{1},\ldots ,t_{m}\}\in
F)\{\prod_{j=1}^{m}\lambda (t_{j})\}dt_{1}\cdots dt_{m},
\end{eqnarray*}%
and, more generally, for any function $h:\mathcal{N}\rightarrow \lbrack
0,\infty )$ 
\begin{equation}
E\{h(X_{B})\}=\sum_{m=0}^{\infty }\int_{B}\cdots \int_{B}h(\{t_{1},\ldots
,t_{m}\})f(\{t_{1},\ldots ,t_{m}\})dt_{1}\cdots dt_{m}.  \label{eq:Eh}
\end{equation}%
A function $h$ on $\mathcal{N}$ is, essentially, a function well defined on $%
\mathcal{S}^{m}$ for any integer $m$ which is invariant under permutation of
the coordinates; for example, $h(\{t_{1},\ldots
,t_{m}\})=\sum_{j=1}^{m}t_{j}/m$.

Single realizations of point processes are often modeled as Poisson
processes with fixed $\lambda $s. But for replicated point processes a
single intensity function $\lambda $ rarely provides an adequate fit for all
replications. It is more reasonable to assume that the $\lambda $s are
subject-specific and treat them as latent random effects. These processes
are called doubly stochastic or Cox processes (M\o ller and Waagepetersen,
2004, ch.~5; Streit, 2010, ch.~8). We can think of a doubly stochastic
process as a pair $(X,\Lambda )$ where $X|\Lambda =\lambda $ is a Poisson
process with intensity $\lambda $, and $\Lambda $ is a random function that
takes values on the space $\mathcal{F}$ of non-negative locally integrable
functions on $\mathcal{S}$. This provides an appropriate framework for the
type of applications we have in mind: the $n$ replications can be seen as
i.i.d.~realizations $(X_{1},\Lambda _{1}),\ldots ,(X_{n},\Lambda _{n})$ of a
doubly stochastic process $(X,\Lambda )$, where $X$ is observable but $%
\Lambda $ is not. In this paper we will assume that all $X_{i}$s are
observed on a common region $B$ of $\mathcal{S}$; the method could be
extended to $X_{i}$s observed on non-conformal regions $B_{i}$ at the
expense of a higher computational complexity.

For many applications we can think of the intensity process $\Lambda $ as a
finite combination of independent components, 
\begin{equation}
\Lambda (t)=\sum_{k=1}^{p}U_{k}\phi _{k}(t),  \label{eq:Lambda_exp}
\end{equation}%
where $\phi _{1},\ldots ,\phi _{p}$ are functions in $\mathcal{F}$ and $%
U_{1},\ldots ,U_{p}$ are independent nonnegative random variables. This is
the functional equivalent of the multivariate Independent Component
decomposition (Hyv\"{a}rinen et al.,~2001). For identifiability we assume
that $\int_{B}\phi _{k}(t)dt=1$ for all $k$, i.e.~that the $\phi _{k}$s are
density functions. Then $\int_{B}\Lambda (t)dt=\sum_{k=1}^{p}U_{k}$ and, if
we define $S=\sum_{k=1}^{p}U_{k}$ and $W_{k}=U_{k}/S$, we have 
\begin{equation}
\Lambda (t)=S\tilde{\Lambda}(t)\text{, with }\tilde{\Lambda}%
(t)=\sum_{k=1}^{p}W_{k}\phi _{k}(t)\text{.}  \label{eq:Lambda_exp_2}
\end{equation}%
The intensity process $\Lambda (t)$, then, can be seen as the product of an
intensity factor $S$ and a scatter factor $\tilde{\Lambda}(t)$ which is a
convex combination of the $\phi _{k}$s. Conditionally on $\Lambda
(t)=\lambda (t)$ the count function $N(B)$ has a Poisson distribution with
rate $S=s$, and conditionally on $N(B)=m$ the $m$ points in $X_{B}$ are
independent identically distributed realizations~with density $\tilde{\lambda%
}(t)$ which is determined by the $W_{k}$s. In general $S$ and $\mathbf{W}$
are not independent, but if the $U_{k}$s are assumed to follow independent
Gamma distributions, say $U_{k}\sim \mathrm{Gamma}(\alpha _{k},\beta )$ with
a common $\beta $, then $S$ and $\mathbf{W}$ are independent with $S\sim 
\mathrm{Gamma}(\sum \alpha _{k},\beta )$ and $\mathbf{W}\sim \mathrm{%
Dirichlet}(\alpha _{1},\ldots ,\alpha _{p})$ (Bilodeau and Brenner, 1999,
ch.~3).

The marginal density of $X_{B}$, with a slight abuse of notation in writing $%
x_{B}=(m,\mathbf{t})$, can be expressed as 
\begin{eqnarray}
f(x_{B}) &=&\int \int f(m,\mathbf{t},s,\mathbf{w})~ds~d\mathbf{w}
\label{eq:marg_XB} \\
&=&\int \int f(\mathbf{t}\mid m,\mathbf{w})f(m\mid s)f(s,\mathbf{w})~ds~d%
\mathbf{w}  \nonumber
\end{eqnarray}%
with 
\[
f(\mathbf{t}\mid m,\mathbf{w})=\prod_{j=1}^{m}\sum_{k=1}^{p}w_{k}\phi
_{k}(t_{j}). 
\]%
Since $\tilde{\lambda}(t)$ is a mixture of the $\phi _{k}$s, for
computational convenience we introduce indicator vectors $\mathbf{y}%
_{1},\ldots ,\mathbf{y}_{m}$ such that, for each $j$, $y_{jk}=1$ for some $k$
and $y_{jk^{\prime }}=0$ for all $k^{\prime }\neq k$, with $P(Y_{jk}=1\mid m,%
\mathbf{w})=w_{k}$. In words, $\mathbf{y}_{j}$ indicates which $\phi _{k}$
the point $t_{j}$ is coming from. Conditionally on $m$ and $\mathbf{w}$, the 
$\mathbf{y}_{j}$s are independent $\mathrm{Multinomial}(1,\mathbf{w}).$
Then, collecting all $\mathbf{y}_{j}$s into a single $\mathbf{y}$ for
notational simplicity, (\ref{eq:marg_XB}) can also be written as 
\begin{eqnarray*}
f(x_{B}) &=&\int \int \sum_{\mathbf{y}}f(m,\mathbf{t},\mathbf{y},s,\mathbf{w}%
)~ds~d\mathbf{w} \\
&=&\int \int \sum_{\mathbf{y}}f(\mathbf{t}\mid m,\mathbf{y})f(\mathbf{y}\mid
m,\mathbf{w})f(m\mid s)f(s,\mathbf{w})~ds~d\mathbf{w}
\end{eqnarray*}%
with 
\[
f\left( \mathbf{t}\mid m,\mathbf{y}\right)
=\prod_{j=1}^{m}\prod_{k=1}^{p}\phi _{k}(t_{j})^{y_{jk}} 
\]%
and 
\[
f(\mathbf{y}\mid m,\mathbf{w})=\prod_{j=1}^{m}\prod_{k=1}^{p}w_{k}^{y_{jk}}. 
\]%
Although the $\mathbf{y}_{j}$s may at first look like an unnecessary
complication, they actually simplify the EM algorithm and are also useful
for data analysis, as we will show in Section \ref{sec:Example}.

To estimate the $\phi _{k}$s we use a semiparametric approach: we assume
they belong to a family of functions $\mathcal{B}$ with non-negative basis
functions $\beta _{1}(t),\ldots ,\beta _{q}(t)$ defined on the region of
interest $B$; for example, we can use B-splines for temporal processes and
tensor-product splines or radial basis functions for spatial processes. Then 
$\phi _{k}(t)=\mathbf{c}_{k}^{T}\mathbf{\beta }(t)$ with $\mathbf{c}_{k}\geq
0$ to guarantee that $\phi _{k}(x)\geq 0$. Note that the specific nature of $%
\mathcal{S}$, i.e.~whether the process is temporal or spatial, only plays a
role here, in the specification of $\mathcal{B}$; for all other purposes we
make no distinctions between temporal and spatial processes.

The model parameters are estimated by penalized maximum likelihood. In
addition to the component coefficients $\mathbf{c}_{1},\ldots ,\mathbf{c}%
_{p} $, the distribution of $S$ and $\mathbf{W}$ will depend on certain
parameters that we will collect in a vector $\mathbf{\eta }$. The whole
model, then, is parameterized by $\mathbf{\theta }=(\mathbf{c}_{1},\ldots ,%
\mathbf{c}_{p},\mathbf{\eta })$ and the parameter space is $\Theta =\mathcal{%
C}\times \mathcal{H}$, where 
\[
\mathcal{C}=\left\{ (\mathbf{c}_{1},\ldots ,\mathbf{c}_{p}):\mathbf{a}^{T}%
\mathbf{c}_{k}=1,\ \mathbf{c}_{k}\geq 0\ \text{for all }k\right\} , 
\]%
with $\mathbf{a}=\int_{B}\mathbf{\beta }(t)dt$, and $\mathcal{H}$\ is the
space of the $\mathbf{\eta }$s. Then, given $n$ independent realizations $%
x_{B1},\ldots ,x_{Bn}$ with $x_{Bi}=\{t_{i1},\ldots ,t_{im_{i}}\}$, the
likelihood function is 
\begin{equation}
L(\mathbf{\theta })=\prod_{i=1}^{n}f(x_{Bi};\mathbf{\theta })
\label{eq:Data_lik}
\end{equation}%
with $f(x_{B};\mathbf{\theta })$ given by (\ref{eq:marg_XB}). Since the
dimension of $\mathcal{B}$ is typically large, a roughness penalty has to be
added to (\ref{eq:Data_lik}) to obtain smooth $\hat{\phi}_{k}$s. We use a
penalty of the form $-\zeta P(\mathbf{\theta })$ with $\zeta \geq 0$ and $P(%
\mathbf{\theta })=\sum_{k=1}^{p}g(\phi _{k})$, where 
\[
g(\phi )=\int_{B}\left\vert \mathrm{H}\phi (t)\right\vert _{F}^{2}\ dt, 
\]%
$\mathrm{H}$ denotes the Hessian and $\left\vert \cdot \right\vert _{F}$ the
Frobenius matrix norm. So for a temporal process $g(\phi )=\int (\phi
^{\prime \prime })^{2}$ and for a spatial process $g(\phi )=\int \{(\frac{%
\partial ^{2}\phi }{\partial t_{1}^{2}})^{2}+2(\frac{\partial ^{2}\phi }{%
\partial t_{1}\partial t_{2}})^{2}+(\frac{\partial ^{2}\phi }{\partial
t_{2}^{2}})^{2}\}$, both of which are quadratic functions in the
coefficients $\mathbf{c}_{k}$. Then $\mathbf{\hat{\theta}}$ maximizes 
\begin{equation}
\rho _{n}(\mathbf{\theta })=n^{-1}\log L(\mathbf{\theta })-\zeta P(\mathbf{%
\theta })  \label{eq:Pen-log-lik}
\end{equation}%
among $\mathbf{\theta }\in \Theta $ for a given smoothing parameter $\zeta $%
. The smoothing parameter $\zeta $ as well as the number of components $p$
can be chosen by cross-validation, maximizing 
\begin{equation}
\func{CV}(\zeta ,p)=\sum_{i=1}^{n}\log f(x_{Bi};\mathbf{\hat{\theta}}%
_{(-i)}),  \label{eq:cv_crit}
\end{equation}%
where $\mathbf{\hat{\theta}}_{(-i)}$ denotes the penalized maximum
likelihood estimator for the reduced sample with $x_{Bi}$ deleted. An EM
algorithm for the computation of $\mathbf{\hat{\theta}}$ is described in
detail in the Supplementary Material.

\section{Asymptotics\label{sec:Asymptotics}}

The asymptotic behavior of $\mathbf{\hat{\theta}}$ as $n\rightarrow \infty $
can be studied via standard empirical-process techniques (Pollard, 1984; Van
der Vaart, 2000), since (\ref{eq:Pen-log-lik}) is the average of independent
identically distributed functions plus a non-random roughness penalty, as in
e.g.~Knight and Fu (2000).

In principle two types of asymptotics may be of interest: the
\textquotedblleft nonparametric\textquotedblright\ asymptotics, where the
dimension of the basis space $\mathcal{B}$ grows with $n$, and the
\textquotedblleft parametric\textquotedblright\ asymptotics, where the true $%
\phi _{k}$s are assumed to belong to $\mathcal{B}$ and then the dimension of 
$\mathcal{B}$ is held fixed. We will follow the second approach here, which
is simpler and sufficient for practical purposes and has been followed by
others (e.g.~Yu and Ruppert, 2002, and Xun et al., 2013) in similar
semiparametric contexts.

The constraints of the space $\mathcal{C}$ introduce some complications in
an otherwise standard asymptotic analysis. We will follow the approach of
Geyer (1994). Proofs of the results presented here can be found in the
Supplementary Material. The first result of this section, consistency of the
estimator, is essentially a consequence of model identifiability. It is not
obvious that model (\ref{eq:Lambda_exp}) is identifiable, so this is proved
first in the Supplementary Material.

\begin{theorem}
\label{thm:Constcy}If the smoothing parameter $\zeta =\zeta _{n}$ goes to
zero as $n\rightarrow \infty $, either deterministically or in probability,
then $\mathbf{\hat{\theta}}_{n}%
\overset{P}{\rightarrow}%
\mathbf{\theta }_{0}.$
\end{theorem}

As explained in Section \ref{sec:Model}, the parameter space $\Theta $ is of
the form $\mathcal{C}\times \mathcal{H}$\ with an $\mathcal{H}$ that we will
assume convex and open. For example, for the model with independent Gamma
scores, $\mathbf{\eta }=(\alpha _{1},\ldots ,\alpha _{p},\beta )$ and $%
\mathcal{H}=(0,+\infty )^{p+1}$. So $\mathcal{H}$ is not a problem for the
asymptotics because it does not contribute active constraints. The problem
is the space $\mathcal{C}$, which is a convex but closed set. In particular,
the non-negativity constraints create some unusual asymptotics; for example,
if a given $c_{0,kj}$ is zero, it is clear that $\sqrt{n}(\hat{c}%
_{n,kj}-c_{0,kj})$ cannot be asymptotically normal since it is a nonnegative
quantity for all $n$.

To handle these constraints we need to introduce the notion of tangent cone.
Let $r$ and $d$ be such that $\mathcal{H}\subseteq \mathbb{R}^{r}$ and $%
\Theta \subseteq \mathbb{R}^{d}$. The tangent cone of $\Theta $ at $\mathbf{%
\theta }_{0}$ is the set of all directions and limits of directions from
which $\mathbf{\theta }_{0}$ can be approached with sequences that stay
inside $\Theta $. If $\mathbf{\theta }_{0}$ is in the interior of $\Theta $
it can be approached from any direction, so the tangent cone is the whole $%
\mathbb{R}^{d}$ in that case. But if $\mathbf{\theta }_{0}$ is on the border
of $\Theta $ it can only be approached from certain directions; for example,
if $c_{0,11}=0$ then the approaching sequences must necessarily satisfy $%
c_{11}\geq 0$ to stay inside $\Theta $. An in-depth treatment of tanget
cones can be found in Hiriart-Urruty and Lemar\'{e}chal (2001, ch.~A.5). The
tangent cone of the product of convex sets is the product of the respective
tangent cones, so the tangent cone of $\Theta $ at $\mathbf{\theta }_{0}$ is 
$\Delta _{0}=\mathcal{T}\times \mathbb{R}^{r}$, with 
\begin{equation}
\mathcal{T}=\left\{ (\mathbf{v}_{1},\ldots ,\mathbf{v}_{p}):\mathbf{a}^{T}%
\mathbf{v}_{k}=0,\ v_{kj}\geq 0\ \text{for }j\in I_{k}\text{, }k=1,\ldots
,p\right\} ,  \label{eq:T_cone}
\end{equation}%
and $I_{k}=\left\{ j:c_{0,kj}=0\right\} $.

The following theorem gives the asymptotic distribution of $\mathbf{\hat{%
\theta}}_{n}$ in a form as explicit as can be given for a general $\mathbf{%
\theta }_{0}$. Let $\mathbf{F}_{0}$ be Fisher's Information Matrix, 
\begin{eqnarray*}
\mathbf{F}_{0} &=&E_{\mathbf{\theta }_{0}}\{\nabla \log f(X_{B};\mathbf{%
\theta }_{0})\nabla \log f(X_{B};\mathbf{\theta }_{0})^{T}\} \\
&=&-E_{\mathbf{\theta }_{0}}\{\nabla ^{2}\log f(X_{B};\mathbf{\theta }%
_{0})\},
\end{eqnarray*}%
where the derivatives of $\log f(x_{B};\mathbf{\theta })$ are taken with
respect to the parameter $\mathbf{\theta }$.

\begin{theorem}
\label{thm:Asymp}If $\sqrt{n}\zeta _{n}\rightarrow \kappa $ as $n\rightarrow
\infty $, either deterministically or in probability, then $\sqrt{n}(\mathbf{%
\hat{\theta}}_{n}-\mathbf{\theta }_{0})%
\overset{D}{\rightarrow}%
\mathbf{\delta }(\mathbf{Z})$, where $\mathbf{\delta }(\mathbf{Z})$ is the
maximizer of 
\[
W(\mathbf{\delta })=\{\mathbf{Z}-\kappa \nabla P(\mathbf{\theta }_{0})\}^{T}%
\mathbf{\delta }-\frac{1}{2}\mathbf{\delta }^{T}\mathbf{F}_{0}\mathbf{\delta 
} 
\]%
over $\mathbf{\delta }\in \Delta _{0}$, and $\mathbf{Z}\sim \mathrm{N}(%
\mathbf{0},\mathbf{F}_{0})$.
\end{theorem}

Although a closed expression for $\mathbf{\delta }(\mathbf{Z})$ in Theorem %
\ref{thm:Asymp} is not available in general, $\mathbf{\delta }(\mathbf{Z})$
is easy to simulate: one generates $M$ Monte Carlo samples $\mathbf{Z}%
_{1},\ldots ,\mathbf{Z}_{M}$ from a $\mathrm{N}(\mathbf{0},\mathbf{F}_{0})$
distribution and then computes $\mathbf{\delta }(\mathbf{Z}_{i})$ for each $%
\mathbf{Z}_{i}$, which involves a simple quadratic optimization problem with
linear constraints. But in the particular case where all component
coefficients $c_{0,kj}$ are strictly positive, we can be more explicit:
write a $\mathbf{\delta }\in \Delta _{0}$ as $\mathbf{\delta }=(\mathbf{%
\delta }_{1},\mathbf{\delta }_{2})$ with $\mathbf{\delta }_{1}\in \mathcal{T}
$ and $\mathbf{\delta }_{2}\in \mathbb{R}^{r}$; since there are no
inequality constraints in (\ref{eq:T_cone}), the only restriction for $%
\mathbf{\delta }_{1}$ is that $(I_{p}\otimes \mathbf{a}^{T})\mathbf{\delta }%
_{1}=\mathbf{0}_{p}$. The matrix $I_{p}\otimes \mathbf{a}^{T}$ is $p\times
pq $ of rank $p$, so there exists an orthogonal $pq\times (pq-p)$ matrix $%
\mathbf{\Gamma }$ such that $(I_{p}\otimes \mathbf{a}^{T})\mathbf{\Gamma }=%
\mathbf{O}$ and $\mathbf{\delta }_{1}=\mathbf{\Gamma \xi }$ for some
unconstrained $\mathbf{\xi }\in \mathbb{R}^{pq-p}$. Let 
\[
\mathbf{A}=\left( 
\begin{array}{cc}
\mathbf{\Gamma }^{T} & \mathbf{O} \\ 
\mathbf{O} & \mathbf{I}_{r}%
\end{array}%
\right) . 
\]
Then 
\[
W(\mathbf{\delta })=\mathbf{\tilde{Z}}^{T}\left( 
\begin{array}{c}
\mathbf{\xi } \\ 
\mathbf{\delta }_{2}%
\end{array}%
\right) -\frac{1}{2}\left( \mathbf{\xi }^{T},\mathbf{\delta }_{2}^{T}\right) 
\mathbf{\tilde{F}}_{0}\left( 
\begin{array}{c}
\mathbf{\xi } \\ 
\mathbf{\delta }_{2}%
\end{array}%
\right) 
\]%
with $\mathbf{\tilde{Z}}=\mathbf{A}\{\mathbf{Z}-\kappa \nabla P(\mathbf{%
\theta }_{0})\}$ and $\mathbf{\tilde{F}}_{0}=\mathbf{AF}_{0}\mathbf{A}^{T}$,
so 
\[
\mathbf{\delta }(\mathbf{Z})=\mathbf{A}\left( 
\begin{array}{c}
\mathbf{\xi }(\mathbf{\tilde{Z})} \\ 
\mathbf{\delta }_{2}(\mathbf{\tilde{Z})}%
\end{array}%
\right) \text{ with }\left( 
\begin{array}{c}
\mathbf{\xi }(\mathbf{\tilde{Z})} \\ 
\mathbf{\delta }_{2}(\mathbf{\tilde{Z})}%
\end{array}%
\right) =\mathbf{\tilde{F}}_{0}^{-1}\mathbf{\tilde{Z}.} 
\]%
Note that $\mathbf{\tilde{Z}}\sim \mathrm{N}(\mathbf{\mu },\mathbf{\tilde{F}}%
_{0})$ with $\mathbf{\mu }=-\kappa \mathbf{A}\nabla P(\mathbf{\theta }_{0})$%
, so $(\mathbf{\xi }(\mathbf{\tilde{Z}),\delta }_{2}(\mathbf{\tilde{Z})})$
has a non-singular $\mathrm{N}(\mathbf{\tilde{F}}_{0}^{-1}\mathbf{\mu },%
\mathbf{\tilde{F}}_{0}^{-1})$ distribution but $\mathbf{\delta }(\mathbf{Z})$
itself has a singular $\mathrm{N}(\mathbf{A\tilde{F}}_{0}^{-1}\mathbf{\mu },%
\mathbf{A\tilde{F}}_{0}^{-1}\mathbf{A}^{T})$ distribution. Nevertheless, the
diagonal elements of $\mathbf{A\tilde{F}}_{0}^{-1}\mathbf{A}^{T}/n$ can be
used as variance estimators to construct, for example, confidence intervals
for the elements of $\mathbf{\theta }$.

\section{Simulations\label{sec:Simulations}}

We ran some simulations to study the finite-sample performance of the
estimators. We considered two models like (\ref{eq:Lambda_exp}), both with
two components: Model 1 had $\phi _{1}(t)=\exp \{-100(t-.3)^{2}\}/.177$ and $%
\phi _{2}(t)=\exp \{-100(t-.7)^{2}\}/.177$ as components, for $t\in \lbrack
0,1]$, which are two peaks with practically no overlap; Model 2 had $\phi
_{1}(t)=\exp \{-20(t-.3)^{2}\}/.385$ and $\phi _{2}(t)=\exp
\{-20(t-.7)^{2}\}/.385$, for $t\in \lbrack 0,1]$, which are also unimodal
but flatter functions with more overlap. The component scores $U_{1}$ and $%
U_{2}$ had lognormal distributions with $E(U_{1})=30$, $E(U_{2})=20$, $%
V(U_{1})=10$ and $V(U_{2})=1$ for both models. Two sample sizes were
considered, $n=50$ and $n=150$.

For estimation we used cubic $B$-splines for the $\phi _{k}$s, with $K=5$
and $K=10$ equispaced knots, and Gamma-distributed component scores. The
estimating models are somewhat different from the data-generating models,
which allows us to investigate the robustness of the estimators to model
misspecification. The smoothing parameter $\zeta $ was chosen by five-fold
cross-validation; for comparison we also computed estimators with the
optimal $\zeta _{\mathrm{opt}}$ that minimizes the mean squared error, which
gives us a lower bound that cannot be attained in practice but serves as a
benchmark to judge the adequacy of five-fold cross-validation as a method
for choosing $\zeta $.

\begin{table}[tbp] \centering%
\begin{tabular}{llllllllllllllll}
& \multicolumn{15}{c}{Model 1} \\ 
& \multicolumn{7}{c}{$n=50$} &  & \multicolumn{7}{c}{$n=150$} \\ 
& \multicolumn{3}{c}{$K=5$} &  & \multicolumn{3}{c}{$K=10$} &  & 
\multicolumn{3}{c}{$K=5$} &  & \multicolumn{3}{c}{$K=10$} \\ 
Estimator & bias & std & rmse &  & bias & std & rmse &  & bias & std & rmse
&  & bias & std & rmse \\ 
$\phi _{1}$ cv & \multicolumn{1}{c}{.73} & \multicolumn{1}{c}{.07} & 
\multicolumn{1}{c}{.73} & \multicolumn{1}{c}{} & \multicolumn{1}{c}{.30} & 
\multicolumn{1}{c}{.14} & \multicolumn{1}{c}{.33} & \multicolumn{1}{c}{} & 
\multicolumn{1}{c}{.73} & \multicolumn{1}{c}{.09} & \multicolumn{1}{c}{.73}
& \multicolumn{1}{c}{} & \multicolumn{1}{c}{.17} & \multicolumn{1}{c}{.12} & 
\multicolumn{1}{c}{.20} \\ 
$\phi _{2}$ cv & \multicolumn{1}{c}{.81} & \multicolumn{1}{c}{.08} & 
\multicolumn{1}{c}{.82} & \multicolumn{1}{c}{} & \multicolumn{1}{c}{.35} & 
\multicolumn{1}{c}{.17} & \multicolumn{1}{c}{.39} & \multicolumn{1}{c}{} & 
\multicolumn{1}{c}{.82} & \multicolumn{1}{c}{.09} & \multicolumn{1}{c}{.82}
& \multicolumn{1}{c}{} & \multicolumn{1}{c}{.19} & \multicolumn{1}{c}{.14} & 
\multicolumn{1}{c}{.24} \\ 
& \multicolumn{1}{c}{} & \multicolumn{1}{c}{} & \multicolumn{1}{c}{} & 
\multicolumn{1}{c}{} & \multicolumn{1}{c}{} & \multicolumn{1}{c}{} & 
\multicolumn{1}{c}{} & \multicolumn{1}{c}{} & \multicolumn{1}{c}{} & 
\multicolumn{1}{c}{} & \multicolumn{1}{c}{} & \multicolumn{1}{c}{} & 
\multicolumn{1}{c}{} & \multicolumn{1}{c}{} & \multicolumn{1}{c}{} \\ 
$\phi _{1}$ opt & \multicolumn{1}{c}{.73} & \multicolumn{1}{c}{.04} & 
\multicolumn{1}{c}{.73} & \multicolumn{1}{c}{} & \multicolumn{1}{c}{.18} & 
\multicolumn{1}{c}{.07} & \multicolumn{1}{c}{.19} & \multicolumn{1}{c}{} & 
\multicolumn{1}{c}{.73} & \multicolumn{1}{c}{.03} & \multicolumn{1}{c}{.73}
& \multicolumn{1}{c}{} & \multicolumn{1}{c}{.14} & \multicolumn{1}{c}{.07} & 
\multicolumn{1}{c}{.16} \\ 
$\phi _{2}$ opt & \multicolumn{1}{c}{.81} & \multicolumn{1}{c}{.04} & 
\multicolumn{1}{c}{.81} & \multicolumn{1}{c}{} & \multicolumn{1}{c}{.20} & 
\multicolumn{1}{c}{.09} & \multicolumn{1}{c}{.22} & \multicolumn{1}{c}{} & 
\multicolumn{1}{c}{.82} & \multicolumn{1}{c}{.03} & \multicolumn{1}{c}{.82}
& \multicolumn{1}{c}{} & \multicolumn{1}{c}{.15} & \multicolumn{1}{c}{.08} & 
\multicolumn{1}{c}{.17} \\ 
&  &  &  &  &  &  &  &  &  &  &  &  &  &  &  \\ 
& \multicolumn{15}{c}{Model 2} \\ 
& bias & std & rmse &  & bias & std & rmse &  & bias & std & rmse &  & bias
& std & rmse \\ 
$\phi _{1}$ cv & \multicolumn{1}{c}{.31} & \multicolumn{1}{c}{.07} & 
\multicolumn{1}{c}{.32} & \multicolumn{1}{c}{} & \multicolumn{1}{c}{.19} & 
\multicolumn{1}{c}{.08} & \multicolumn{1}{c}{.20} & \multicolumn{1}{c}{} & 
\multicolumn{1}{c}{.31} & \multicolumn{1}{c}{.05} & \multicolumn{1}{c}{.32}
& \multicolumn{1}{c}{} & \multicolumn{1}{c}{.19} & \multicolumn{1}{c}{.04} & 
\multicolumn{1}{c}{.19} \\ 
$\phi _{2}$ cv & \multicolumn{1}{c}{.33} & \multicolumn{1}{c}{.08} & 
\multicolumn{1}{c}{.34} & \multicolumn{1}{c}{} & \multicolumn{1}{c}{.22} & 
\multicolumn{1}{c}{.10} & \multicolumn{1}{c}{.24} & \multicolumn{1}{c}{} & 
\multicolumn{1}{c}{.33} & \multicolumn{1}{c}{.07} & \multicolumn{1}{c}{.34}
& \multicolumn{1}{c}{} & \multicolumn{1}{c}{.22} & \multicolumn{1}{c}{.05} & 
\multicolumn{1}{c}{.23} \\ 
& \multicolumn{1}{c}{} & \multicolumn{1}{c}{} & \multicolumn{1}{c}{} & 
\multicolumn{1}{c}{} & \multicolumn{1}{c}{} & \multicolumn{1}{c}{} & 
\multicolumn{1}{c}{} & \multicolumn{1}{c}{} & \multicolumn{1}{c}{} & 
\multicolumn{1}{c}{} & \multicolumn{1}{c}{} & \multicolumn{1}{c}{} & 
\multicolumn{1}{c}{} & \multicolumn{1}{c}{} & \multicolumn{1}{c}{} \\ 
$\phi _{1}$ opt & \multicolumn{1}{c}{.30} & \multicolumn{1}{c}{.08} & 
\multicolumn{1}{c}{.31} & \multicolumn{1}{c}{} & \multicolumn{1}{c}{.09} & 
\multicolumn{1}{c}{.07} & \multicolumn{1}{c}{.11} & \multicolumn{1}{c}{} & 
\multicolumn{1}{c}{.30} & \multicolumn{1}{c}{.05} & \multicolumn{1}{c}{.30}
& \multicolumn{1}{c}{} & \multicolumn{1}{c}{.08} & \multicolumn{1}{c}{.06} & 
\multicolumn{1}{c}{.10} \\ 
$\phi _{2}$ opt & \multicolumn{1}{c}{.32} & \multicolumn{1}{c}{.07} & 
\multicolumn{1}{c}{.33} & \multicolumn{1}{c}{} & \multicolumn{1}{c}{.07} & 
\multicolumn{1}{c}{.08} & \multicolumn{1}{c}{.11} & \multicolumn{1}{c}{} & 
\multicolumn{1}{c}{.32} & \multicolumn{1}{c}{.04} & \multicolumn{1}{c}{.32}
& \multicolumn{1}{c}{} & \multicolumn{1}{c}{.07} & \multicolumn{1}{c}{.07} & 
\multicolumn{1}{c}{.10}%
\end{tabular}%
\caption{Simulation Results. Bias, standard deviation and root mean squared error of independent
component estimators.}\label{tab:Sim_1}%
\end{table}%

Table \ref{tab:Sim_1} summarizes the results. For each estimator we report $%
\mathrm{bias}=(\int [E\{\hat{\phi}(t)\}-\phi (t)]^{2}dt)^{1/2}$, $\mathrm{std%
}=\{\int E([\hat{\phi}(t)-E\{\hat{\phi}(t)\}]^{2})dt\}^{1/2}$ and $\mathrm{%
rmse}=(\int E[\{\hat{\phi}(t)-\phi (t)\}^{2}]dt)^{1/2}$, where the
expectations have been approximated by 500 Monte Carlo replications. We do
not report Monte Carlo standard errors on the table, but, for example, the
first mean squared error reported, $.732^{2}=.536$, has a Monte Carlo
standard deviation of $.001$, which gives a 95\% confidence interval $%
(.534,.538)$ for the mean squared error and $(.730,.733)$ for the root mean
squared error, so the quantities in Table \ref{tab:Sim_1} are accurate up to
the two reported decimals. We see in Table \ref{tab:Sim_1} that the
comparative behavior of the estimators with respect to sample size and
number of knots is similar for both models and both components, although $%
\phi _{1}$ is easier to estimate than $\phi _{2}$ because it is the dominant
component. We see that ten knots produce much better estimators than five
knots, even for a sample size as small as 50. For a fixed sample size we see
that increasing the number of knots reduces the bias and increases the
variance, as expected, but the gains in bias amply compensate for the
increases in variance, so the overall mean squared errors are much smaller.
The estimators, then, successfully \textquotedblleft borrow
strength\textquotedblright\ across replications, and it is best to err on
the side of using too many basis functions rather than too few. Overall, the
performance of the cross-validated estimators is not much worse than that of
the optimal estimators, so five-fold cross-validation is to be an adequate
method for choosing the smoothing parameter, although there is some room for
improvement.

It is also of interest to compare the behavior of the new method with the
method of Wu et al. (2013). Direct comparison of the component estimators is
not possible because the method of Wu et al. (2013) is not based on a
non-negative decomposition for the process $\Lambda (t)$ like model (\ref%
{eq:Lambda_exp}). Briefly, the method of Wu et al.~(2013) works as follows:
first, $\mu (t)=E\{\tilde{\Lambda}(t)\}$ and $\rho (s,t)=\mathrm{Cov}\{%
\tilde{\Lambda}(s),\tilde{\Lambda}(t)\}$ are estimated by kernel smoothers,
and the eigenfunctions of $\rho $, which can be negative, are computed; then
the individual densities $\tilde{\lambda}_{i}$ are estimated using the
eigenfunctions of $\rho (s,t)$ as basis, truncating and renormalizing the $%
\tilde{\lambda}_{i}$s if necessary to make them non-negative; finally, the
intensities $\lambda _{i}$ are estimated as $\lambda _{i}=m_{i}\tilde{\lambda%
}_{i}$, where $m_{i}$ is the number of observations for replication $i$.
Since we cannot compare the component estimators directly, we will compare
the intensity estimators and the density estimators. The method of Wu et
al.~(2013) has three main tuning parameters: the bandwidths of the kernel
smoothers and the number of components to include in the expansions of the $%
\tilde{\lambda}_{i}$s. Wu et al. (2013) discuss a number of strategies to
choose these parameters. Here we use the optimal Gaussian bandwidths for the
kernel smoothers (Silverman, 1986, ch.~3.4 and 4.3), which produce
reasonably smooth estimators of $\mu (t)$ and $\rho (s,t)$ in our
simulations, and instead of choosing the number of components $p$ using Wu
et al.'s suggestions we simply report the estimation errors for the optimal $%
p$, which happens to correspond to $p=1$. Note that Wu et al.'s model has a
mean while (\ref{eq:Lambda_exp}) does not, so a one-principal-component
model has the same dimension as a two-independent-component model.

\begin{table}[tbp] \centering%
\begin{tabular}{llllll}
& \multicolumn{5}{c}{Model 1} \\ 
& \multicolumn{2}{c}{$n=50$} &  & \multicolumn{2}{c}{$n=150$} \\ 
Estimator & intensity & density &  & intensity & density \\ 
IC-based, cv & \multicolumn{1}{c}{14} & \multicolumn{1}{c}{.26} & 
\multicolumn{1}{c}{} & \multicolumn{1}{c}{11} & \multicolumn{1}{c}{.18} \\ 
IC-based, opt & \multicolumn{1}{c}{10} & \multicolumn{1}{c}{.17} & 
\multicolumn{1}{c}{} & \multicolumn{1}{c}{10} & \multicolumn{1}{c}{.15} \\ 
PC-based, opt & \multicolumn{1}{c}{13} & \multicolumn{1}{c}{.18} & 
\multicolumn{1}{c}{} & \multicolumn{1}{c}{12} & \multicolumn{1}{c}{.14} \\ 
&  &  &  &  &  \\ 
& \multicolumn{5}{c}{Model 2} \\ 
IC-based, cv & \multicolumn{1}{c}{7.0} & \multicolumn{1}{c}{.10} & 
\multicolumn{1}{c}{} & \multicolumn{1}{c}{6.5} & \multicolumn{1}{c}{.09} \\ 
IC-based, opt & \multicolumn{1}{c}{6.8} & \multicolumn{1}{c}{.10} & 
\multicolumn{1}{c}{} & \multicolumn{1}{c}{6.7} & \multicolumn{1}{c}{.10} \\ 
PC-based, opt & \multicolumn{1}{c}{9.6} & \multicolumn{1}{c}{.12} & 
\multicolumn{1}{c}{} & \multicolumn{1}{c}{9.3} & \multicolumn{1}{c}{.11}%
\end{tabular}%
\caption{Simulation Results. Root mean squared errors of intensity and
density function estimators.}\label{tab:Sim_2}%
\end{table}%

Table \ref{tab:Sim_2} shows the estimated root mean squared errors $%
\{E(\sum_{i=1}^{n}\Vert \hat{\lambda}_{i}-\lambda _{i}\Vert ^{2}/n)\}^{1/2}$
and $\{E(\sum_{i=1}^{n}\Vert \widehat{\tilde{\lambda}}_{i}-\tilde{\lambda}%
_{i}\Vert ^{2}/n)\}^{1/2}$, based on 500 Monte Carlo replications. In fact,
the same simulated data and the same independent-component estimators as in
Table \ref{tab:Sim_1} were used. We only report the results for the ten-knot
estimators, since they were uniformly better than the five-knot estimators.
Again, we report the errors of both the cross-validated and the optimal
independent-component-based estimators. Note that for Wu et al.'s
principal-component-based estimators we are choosing the optimal $p$, so the
errors observed in practice will be somewhat higher than those reported in
Table \ref{tab:Sim_2}. Unlike the results in Table \ref{tab:Sim_1}, the
results in Table \ref{tab:Sim_2} do change with the model. For Model 1 the
results are mixed, depending on the sample size and on whether we are
estimating the density function or the intensity function. But for Model 2
the independent-component-based estimators, even the cross-validated ones,
clearly outperform Wu et al.'s principal-component-based estimators.

\section{\label{sec:Example}Application: Street Theft in Chicago}

In this section we analyze the spatial distribution of street robberies in
Chicago during the year 2014. The data was downloaded from the City of
Chicago Data Portal, a very extensive data repository that provides, among
other things, detailed information about every reported criminal incident in
the city. The information provided includes type, date, time, and
coordinates (latitude and longitude) of the incident. Here we will focus on
crimes typified as of primary type \textquotedblleft
theft\textquotedblright\ and location \textquotedblleft
street\textquotedblright . There were 16,278 reported incidents of this type
between January 1, 2014 and December 31, 2014. Their locations cover most of
the city, as shown in Figure \ref{fig:Maps}(a).

\FRAME{ftbpFU}{5.9439in}{4.8948in}{0pt}{\Qcb{Chicago Street Theft. (a)
Location of reported incidents in the year 2014. (b) Kernel density
estimator of the data in (a).}}{\Qlb{fig:Maps}}{maps_color.eps}{\special%
{language "Scientific Word";type "GRAPHIC";maintain-aspect-ratio
TRUE;display "ICON";valid_file "F";width 5.9439in;height 4.8948in;depth
0pt;original-width 6.7196in;original-height 4.8672in;cropleft
"0.0596";croptop "1";cropright "0.9402";cropbottom "0";filename
'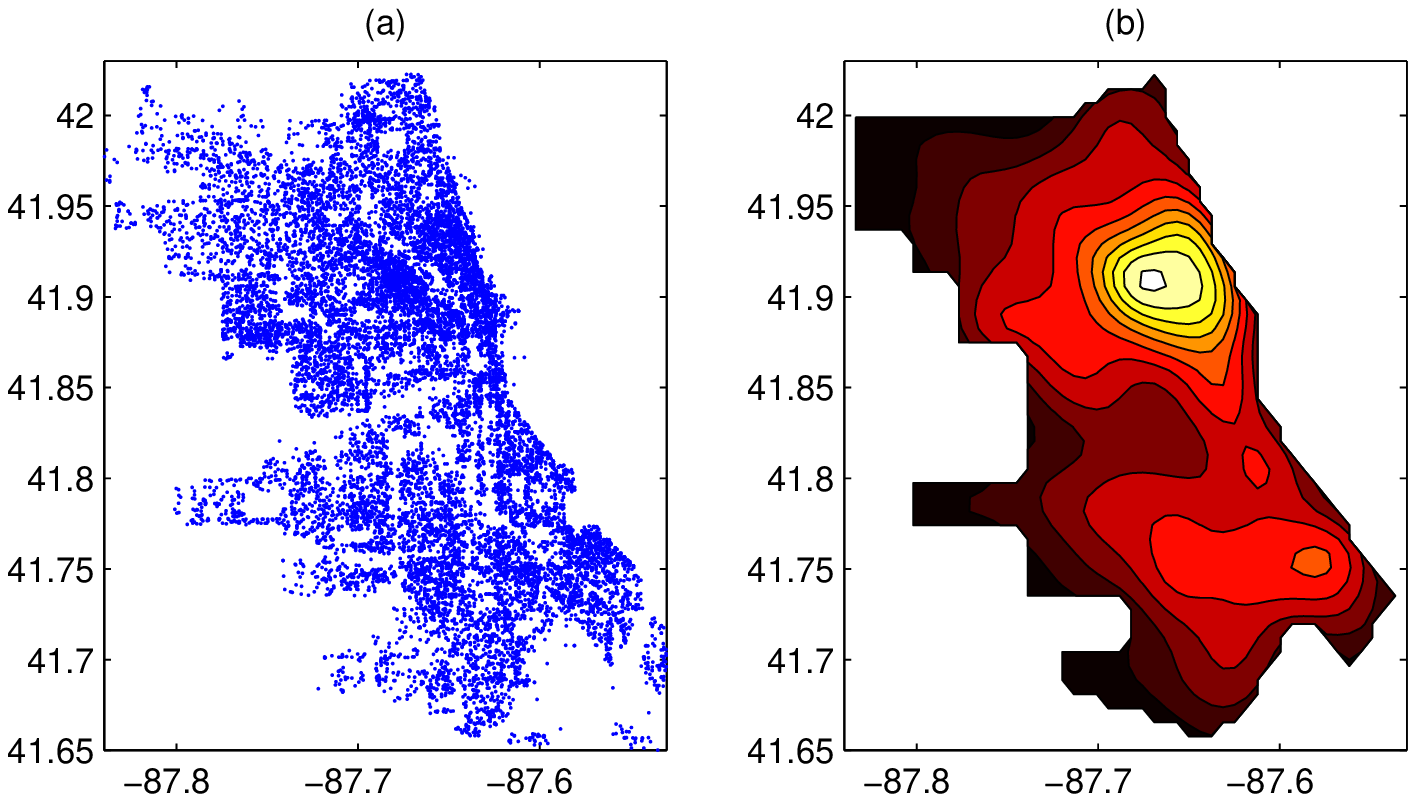';file-properties "XNPEU";}}

Investigating the spatial and temporal distribution of crime is important
because, as noted by Ratcliffe (2010), crime opportunities are not uniformly
distributed in space and time, and the discovery and analysis of spatial
patterns may help better understand the role of geography and opportunity in
the incidence of crime. The geographical analysis of crime is not new and
goes back at least to Guerry and Quetelet in the early 1800s (Friendly,
2007), but technological and data limitations hampered earlier attempts at
crime mapping. Only in the last few decades have software and hardware
developed to the point of allowing efficient crime mapping, and only in the
last few years have extensive crime data repositories like the City of
Chicago Data Portal become available.

Awareness of the geographical and temporal distribution of crime is
necessary, for example, for efficient allocation of crime-prevention
resources. Sometimes even the most basic spatial analysis reveals unexpected
results. Figure \ref{fig:Maps}(b) shows a kernel-density estimator for the
points in Figure \ref{fig:Maps}(a). Those familiar with the Chicago area
will immediately notice that the mode occurs at the neighborhoods of Wicker
Park and Pulaski, an entertainment area with high pedestrian traffic that is
not typically associated with crime and violence in people's mind. But it is
well known to criminologists that street robbery tends to concentrate not on
deprived neighborhoods per se but on places that they denominate
\textquotedblleft crime attractors\textquotedblright , areas that
\textquotedblleft bring together, often in large numbers, people who carry
cash, some of whom are distracted and vulnerable\textquotedblright\
(Bernasco and Block, 2011).

We fitted independent component models for these data, using the 365 days as
replications. We considered $p$s ranging from 2 to 5 and component scores
with Gamma distribution. As basis family $\mathcal{B}$ for the components we
used normalized Gaussian radial kernels with 49 uniformly distributed
centers in the rectangle $[-87.84,-87.53]\times \lbrack 41.65,42.03]$, the
smallest rectangle that includes the domain of interest $B$, the city of
Chicago. The smoothing parameter $\zeta $ was chosen by five-fold
cross-validation for each $p$. The cross-validated criteria for $p=2,\ldots
,5$ were $179.37$, $177.73$, $186.04$ and $185.05$, respectively, so we
chose the model with four components. For this model the $\hat{\alpha}_{k}$s
were $4.95$, $4.38$, $4.04$ and $3.51$, and $\hat{\beta}$ was $2.70$, so the
respective expected values of the $U_{k}$s were $13.37$, $11.85$, $10.93$
and $9.50$. The overall expected number of robberies per day in the city is
then $45.64$.

\FRAME{ftbpFU}{4.5947in}{5.1681in}{0pt}{\Qcb{Chicago Street Theft. Contour
plot of (a) first, (b) second, (c) third, and (d) fourth component.}}{\Qlb{%
fig:Components}}{components_color.eps}{\special{language "Scientific
Word";type "GRAPHIC";maintain-aspect-ratio TRUE;display "USEDEF";valid_file
"F";width 4.5947in;height 5.1681in;depth 0pt;original-width
4.5671in;original-height 5.1404in;cropleft "0";croptop "1";cropright
"1";cropbottom "0";filename '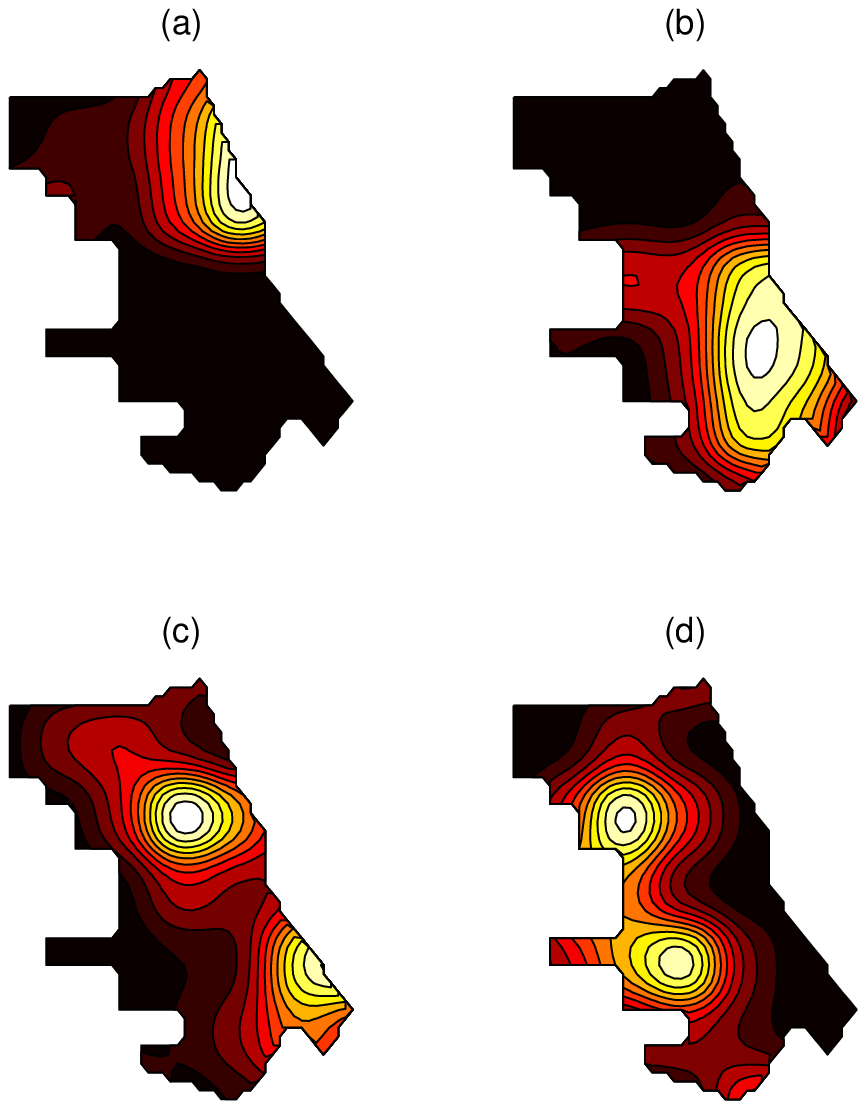';file-properties "XNPEU";}}

Contour plots of the four components are shown in Figure \ref{fig:Components}%
. We computed the asymptotic variance estimators derived in Section \ref%
{sec:Asymptotics} for the component coefficients and it turned out that many 
$\hat{c}_{kj}$s were indeed not significantly different from zero, but the
plots obtained after filtering out these components were virtually identical
to those in Figure \ref{fig:Components}, so they are not shown here.

The components in Figure \ref{fig:Components} are well-localized and easily
interpretable. The first component has a mode on the border of Near North
and Lincoln Park neighborhoods, and extends northwards towards Lakeview and
Uptown. These are the most affluent neighborhoods in Chicago, normally not
associated with crime in people's perceptions, but the large number of
affluent people on the streets make them ideal \textquotedblleft
attractors\textquotedblright\ for street theft. The second component has a
mode on Washington Park and extends eastwards towards the University of
Chicago campus, neighborhoods in the South side of the city that are
well-known to be problematic; the proportion of households below poverty
level in Washington Park is 39\%, and the unemployment level is 23\%.

The third component is bimodal, with the main mode at West Town, roughly
where the mode of the overall mean was (Figure \ref{fig:Components}(b)), and
a second mode at Woodlawn and South Shore neighborhoods. These are also
neighborhoods with generally high levels of crime and poverty, although not
as high as Washington Park. The fourth component is also bimodal. The main
mode concentrates on the West side neighborhoods of Humboldt Park and West
Garfield, and the second mode on the South side neighborhoods of West
Englewood and Gage Park. These are the worst neighboorhoods in the city in
terms of overall crime and poverty; for example, the proportion of
households below poverty level is 33\% in Humboldt Park and 32\% in West
Engelwood, and the percentages of people with no high school diploma are
37\% and 30\%, respectively. These similarities in demographics suggest that
the bimodality of the components is not an artifact of the estimators but a
real feature of the data.

\FRAME{ftbpFU}{5.7536in}{5.3956in}{0pt}{\Qcb{Chicago Street Theft. Incidents
reported in (a) November 8 and (b) July 25 of 2014. Corresponding intensity
estimators based on a four-component model, for (b) November 8 and (d) July
25.}}{\Qlb{fig:Days_312_206}}{days_312_206_color.eps}{\special{language
"Scientific Word";type "GRAPHIC";maintain-aspect-ratio TRUE;display
"USEDEF";valid_file "F";width 5.7536in;height 5.3956in;depth
0pt;original-width 6.429in;original-height 5.3679in;cropleft
"0.1092";croptop "1";cropright "1";cropbottom "0";filename
'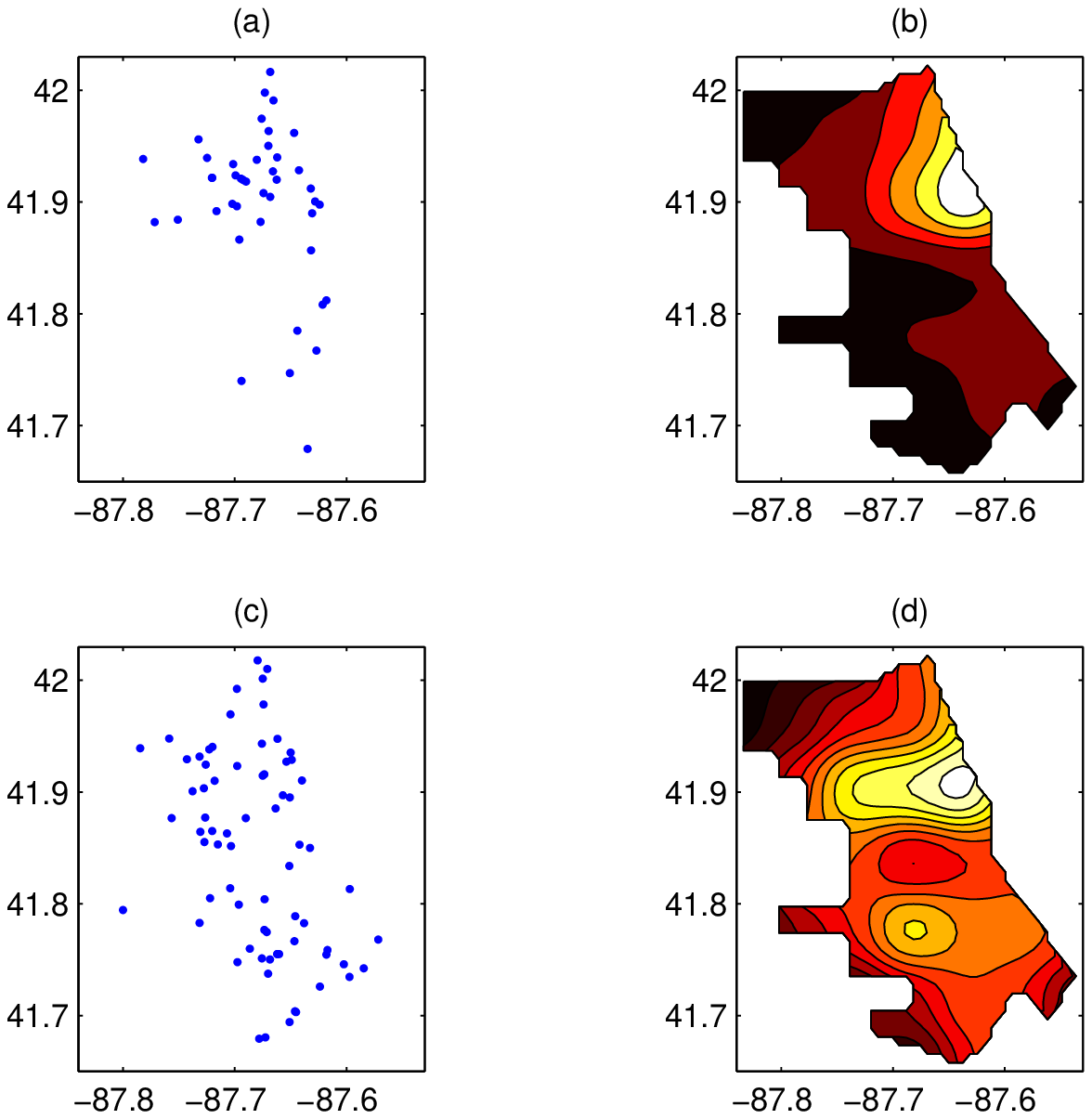';file-properties "XNPEU";}}

Model (\ref{eq:Lambda_exp}) also provides estimators for the individual
intensity functions, $\hat{\lambda}_{i}(t)=\sum_{k=1}^{p}\hat{u}_{ik}\hat{%
\phi}_{k}(t)$. Two examples are shown in Figure \ref{fig:Days_312_206}.
Figure \ref{fig:Days_312_206}(a) corresponds to November 8, the day with the
largest first component score, and Figure \ref{fig:Days_312_206}(c) to July
25, the day with the largest fourth component score. There was a total of 43
thefts in November 8, 35 of which were associated with the first component,
as determined by the $\hat{y}_{ijk}$s; specifically, for each incident $%
t_{ij}$ we found $\hat{k}_{ij}=\limfunc{argmax}_{k}\hat{y}_{ijk}$, which
indicates what component $t_{ij}$ is most strongly associated with. The
corresponding estimator of the intensity function is shown in Figure \ref%
{fig:Days_312_206}(b), which accurately reflects the distribution of the
points in Figure \ref{fig:Days_312_206}(a). In July 25 there was a total of
72 street robberies, 17 of which were associated with the first component,
17 with the second, 0 with the third and 38 with the fourth. The
corresponding intensity function is shown in Figure \ref{fig:Days_312_206}%
(d), and it is again an accurate representation of the distribution of the
points in Figure \ref{fig:Days_312_206}(c). If we were using individual
kernel smoothers for the intensity functions, with only 43 or even 72 data
points we would not be able to obtain estimators that are smooth and
fine-detailed at the same time, like those in Figure \ref{fig:Days_312_206};
a small bandwidth would produce irregular estimators and a large bandwidth
would produce featureless estimators.

\FRAME{ftbpFU}{5.6204in}{4.3716in}{0pt}{\Qcb{Chicago Street Theft. Daily
component scores and lowess smoother (solid line) for (a) first, (b) second,
(c) third, and (d) fourth component.}}{\Qlb{fig:Scores}}{scores_color.eps}{%
\special{language "Scientific Word";type "GRAPHIC";maintain-aspect-ratio
TRUE;display "USEDEF";valid_file "F";width 5.6204in;height 4.3716in;depth
0pt;original-width 5.7943in;original-height 4.3439in;cropleft
"0.0346";croptop "1";cropright "1";cropbottom "0";filename
'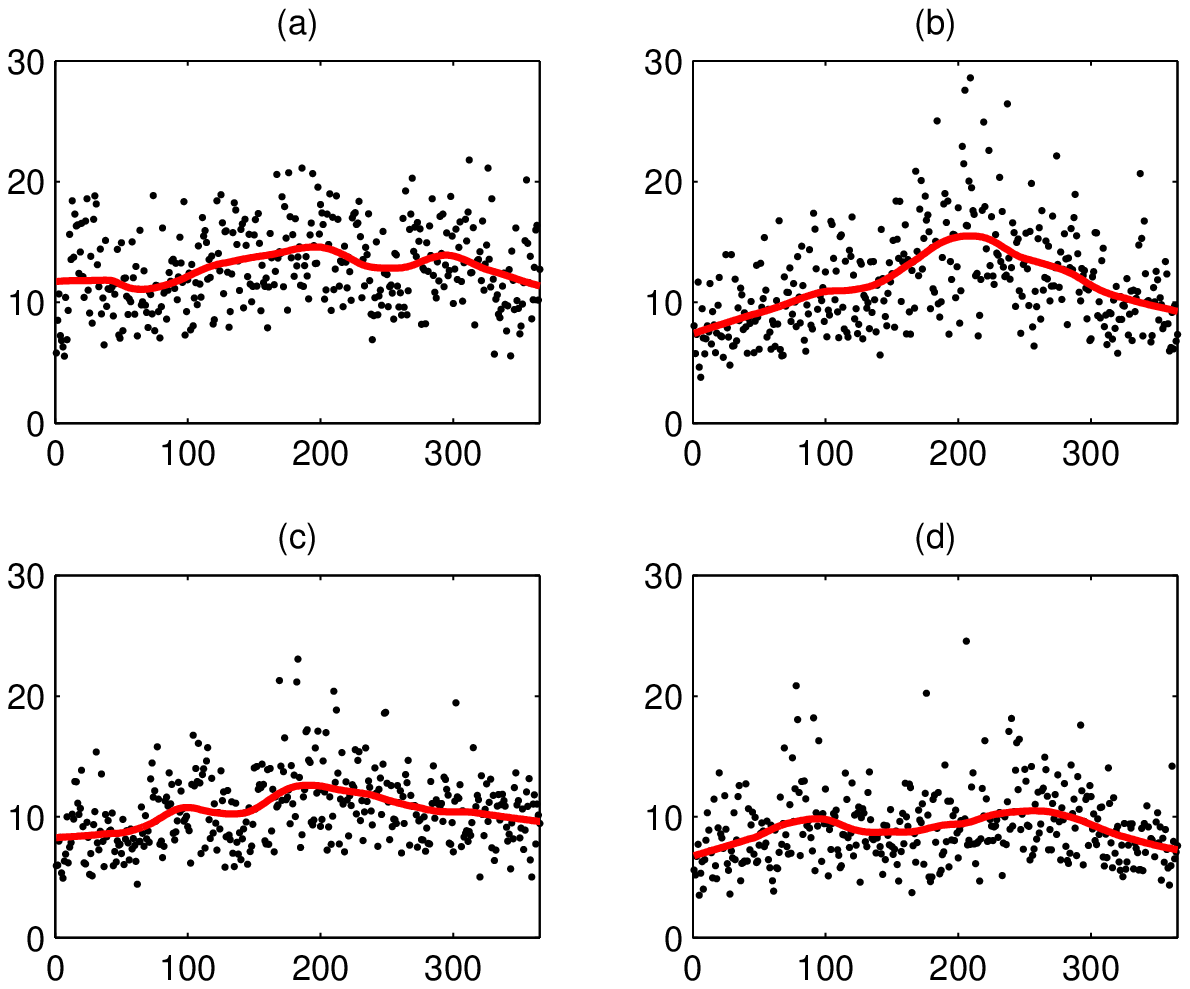';file-properties "XNPEU";}}

An analysis of the component scores also reveals interesting patterns. It is
known that some types of crime tend to follow seasonal patterns; violent
crimes in particular tend to increase during the summer (Field, 1992).
Figure \ref{fig:Scores} shows that this is indeed the case for street theft
in Chicago. Even though there is a large variability, the seasonal trends
are plain to see, in particular for the second component which shows a very
clear peak in July. The first and third components also reach their maxima
in July, although their curves are flatter. The fourth component, on the
other hand, has two local peaks in April and September. Whether these peaks
are systematic across the years may be established by analyzing data from
previous years, since the City of Chicago Data Portal has daily crime data
going back to 2001, but we will not do that here.

\FRAME{ftbpFU}{5.8833in}{4.7011in}{0pt}{\Qcb{Chicago Street Theft. Kernel
density estimators of the distribution of time of reported incidents, for
incidents associated with (a) first, (b) second, (c) third, and (d) fourth
component.}}{\Qlb{fig:Times}}{time.eps}{\special{language "Scientific
Word";type "GRAPHIC";maintain-aspect-ratio TRUE;display "USEDEF";valid_file
"F";width 5.8833in;height 4.7011in;depth 0pt;original-width
7.8222in;original-height 5.8418in;cropleft "0.0319";croptop "1";cropright
"0.9679";cropbottom "0";filename '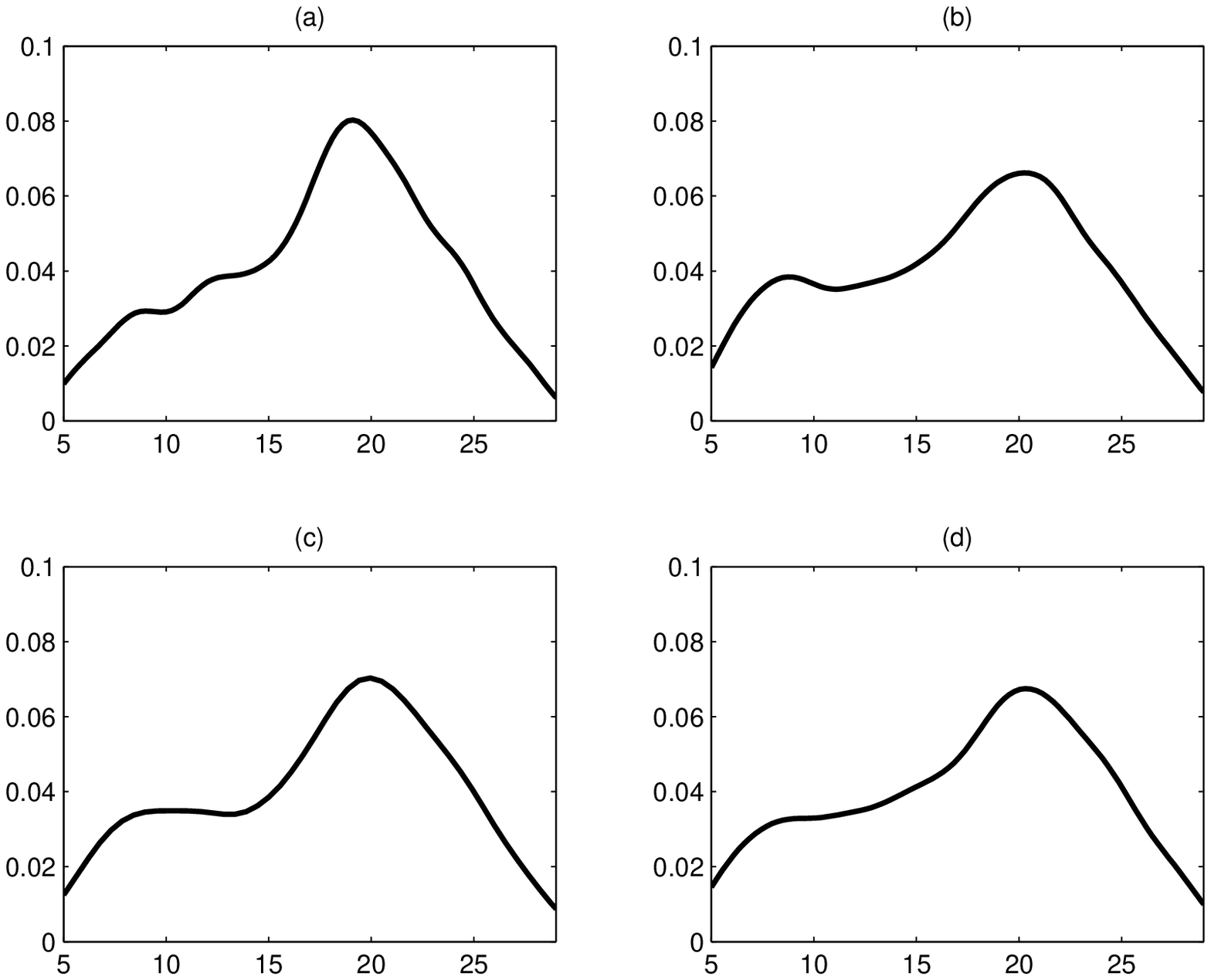';file-properties "XNPEU";}}

Finally, we investigated the distribution of the time of the incidents.
Figure \ref{fig:Times} shows kernel density estimators of the time of
robbery for each component. Since the minimum always occurs at 5am, for
better visualization we shifted the points between 0 and 4am to the other
end by adding 24, so e.g.~3am became 27. We see that the largest number of
incidents tend to occur around 8 pm for all regions, but while the
distribution is strongly unimodal for the North side neighborhoods (Figure %
\ref{fig:Times}(a)), for the other regions there are clear plateaus earlier
in the day (Figures \ref{fig:Times}(b--d)). These plots, however, may hide
seasonal variations like those seen in Figure \ref{fig:Scores}. A more
thorough analysis of these variations would need models that incorporate
covariates, but this is beyond the scope of this paper.

\section*{References}

\begin{description}
\item Baddeley, A.J., Moyeed, R.A., Howard, C.V., and Boyde, A. (1993).
Analysis of a three-dimensional point pattern with replication. \emph{%
Applied Statistics} \textbf{42 }641--668.

\item Bell, M.L., and Grunwald, G.K. (2004). Mixed models for the analysis
of replicated spatial point patterns. \emph{Biostatistics }\textbf{5 }%
633--648.

\item Bernasco, W., and Block, R. (2011). Robberies in Chicago: a
block-level analysis of the influence of crime generators, crime attractors,
and offender anchor points. \emph{Journal of Research in Crime and
Delinquency} \textbf{48} 33--57.

\item Bilodeau, M., and Brenner, D. (1999). \emph{Theory of Multivariate
Statistics.} Springer, New York.

\item Bouzas, P.R., Valderrama, M., Aguilera, A.M., and Ruiz-Fuentes, N.
(2006). Modelling the mean of a doubly stochastic Poisson process by
functional data analysis. \emph{Computational Statistics and Data Analysis} 
\textbf{50} 2655--2667.

\item Bouzas, P.R., Ruiz-Fuentes, N., and Oca\~{n}a, F.M. (2007). Functional
approach to the random mean of a compound Cox process. \emph{Computational
Statistics }\textbf{22} 467--479.

\item Cox, D.R., and Isham, V. (1980). \emph{Point Processes.} Chapman and
Hall/CRC, Boca Raton.

\item Diggle, P.J. (2013). \emph{Statistical Analysis of Spatial and
Spatio-Temporal Point Patterns, Third Edition.} Chapman and Hall/CRC, Boca
Raton.

\item Diggle, P.J., Lange, N., and Bene\v{s}, F.M. (1991). Analysis of
variance for replicated spatial point patterns in clinical neuroanatomy. 
\emph{Journal of the American Statistical Association }\textbf{86} 618--625.

\item Diggle, P.J., Mateau, J., and Clough, H.E. (2000). A comparison
between parametric and nonparametric approaches to the analysis of
replicated spatial point patterns. \emph{Advances in Applied Probability} 
\textbf{32 }331--343.

\item Diggle, P.J., Eglen, S.J., and Troy, J.B. (2006). Modeling the
bivariate spatial distribution of amacrine cells. In \emph{Case Studies in
Spatial Point Process Modeling}, eds.~A. Baddeley et al., New York:
Springer, pp.~215--233.

\item Fern\'{a}ndez-Alcal\'{a}, R.M., Navarro-Moreno, J., and Ruiz-Molina,
J.C. (2012). On the estimation problem for the intensity of a DSMPP. \emph{%
Methodology and Computing in Applied Probability }\textbf{14} 5--16.

\item Field, S. (1992). The effect of temperature on crime. \emph{British
Journal of Criminology }\textbf{32} 340--351.

\item Friendly, M. (2007). A.-M. Guerry's moral statistics of France:
challenges for multivariable spatial analysis. \emph{Statistical Science }%
\textbf{22 }368--399.

\item Hiriart-Urruty, J.-B., and Lemar\'{e}chal, C. (2001). \emph{%
Fundamentals of Convex Analysis.} Springer, NY:

\item Knight, K., and Fu, W. (2000). Asymptotics for lasso-type estimators. 
\emph{The Annals of Statistics }\textbf{28} 1356--1378.

\item Gervini, D. (2009). Detecting and handling outlying trajectories in
irregularly sampled functional datasets. \emph{The Annals of Applied
Statistics} \textbf{3} 1758--1775.

\item Geyer, C.J. (1994). On the asymptotics of constrained M-estimation. 
\emph{The Annals of Statistics }\textbf{22} 1993--2010.

\item Hyv\"{a}rinen, A., Karhunen, J., and Oja, E. (2001). \emph{Independent
Component Analysis.} Wiley, New York.

\item Jalilian, A., Guan, Y., and Waagpetersen, R. (2013). Decomposition of
variance for spatial Cox processes. \emph{Scandinavian Journal of Statistics 
}\textbf{40 }119--137.

\item James, G., Hastie, T., and Sugar, C. (2000). Principal component
models for sparse functional data. \emph{Biometrika} \textbf{87} 587--602.

\item Landau, S., Rabe-Hesketh, S., and Everall, I.P. (2004). Nonparametric
one-way analysis of variance of replicated bivariate spatial point patterns. 
\emph{Biometrical Journal }\textbf{46 }19--34.

\item M\o ller, J., and Waagepetersen, R.P. (2004). \emph{Statistical
Inference and Simulation for Spatial Point Processes}. Chapman and Hall/CRC,
Boca Raton.

\item Pawlas, Z. (2011). Estimation of summary characteristics from
replicated spatial point processes. \emph{Kybernetika} \textbf{47} 880--892.

\item Pollard, D. (1984). \emph{Convergence of Stochastic Processes. }New
York: Springer.

\item Ratcliffe, J. (2010). Crime mapping: spatial and temporal challenges.
In \emph{Handbook of Quantitative Criminology}, A.R. Piquero and D. Weisburd
(eds.), pp.~5--24. New York: Springer.

\item Silverman, B.W. (1986). \emph{Density Estimation for Statistics and
Data Analysis.} Chapman \& Hall.

\item Snyder, D.L., and Miller, M.I. (1991). \emph{Random Point Processes in
Time and Space.} Springer, New York.

\item Streit, R.L. (2010). \emph{Poisson Point Processes: Imaging, Tracking,
and Sensing.} Springer, New York.

\item Van der Vaart, A. (2000). \emph{Asymptotic Statistics}. Cambridge
University Press, Cambridge, UK.

\item Yao, F., M\"{u}ller, H.-G., and Wang, J.-L. (2005). Functional data
analysis for sparse longitudinal data. \emph{Journal of the American
Statistical Association} \textbf{100} 577--590.

\item Wager, C.G., Coull, B.A., and Lange, N. (2004). Modelling spatial
intensity for replicated inhomogeneous point patterns in brain imaging. 
\emph{Journal of the Royal Statistical Society Series B} \textbf{66 }%
429--446.

\item Wu, S., M\"{u}ller, H.-G., and Zhang, Z. (2013). Functional data
analysis for point processes with rare events. \emph{Statistica Sinica }%
\textbf{23} 1--23.

\item Xun, X., Cao, J., Mallick, B., Maity, A., and Carroll, R.J. (2013).
Parameter estimation of partial differential equations. \emph{Journal of the
American Statistical Association} \textbf{108} 1009--1020.

\item Yu, Y., and Ruppert, D. (2002). Penalized spline estimation for
partially linear single-index models. \emph{Journal of the American
Statistical Association} \textbf{97} 1042--1054.
\end{description}

\end{document}